\documentclass{article}

\usepackage{PRIMEarxiv}

\usepackage[utf8]{inputenc} 
\usepackage[T1]{fontenc}    
\usepackage{hyperref}       
\usepackage{url}            
\usepackage{booktabs}       
\usepackage{amsfonts}       
\usepackage{nicefrac}       
\usepackage{microtype}      
\usepackage{lipsum}
\usepackage{fancyhdr}       
\usepackage{graphicx}       
\graphicspath{{media/}}     

\usepackage[T1]{fontenc}
\usepackage{float}
\usepackage{amsmath}
\usepackage{amssymb}

\title{Homeostasis Revisited and Reformulated Through Hidden Markov Model Control}

\author{
  Rubén Moreno-Bote \\
  Department of Engineering and Center for Brain and Cognition \\
  Universitat Pompeu Fabra \\
  08002, Barcelona, Spain\\
  \texttt{ruben.moreno@upf.edu} \\
}

\begin{document}
\maketitle

\begin{abstract}
A common formalization of homeostasis is the free energy principle, a framework that defines a set of desired observation values, or critical states, that the agent should reach or remain close to. 
Under the free energy principle, an agent should act to maximize the probability of receiving the desired observations.
Maximizing this probability pushes the agent to avoid surprising events, those that put it at the risk of homeostatic breakage. 
Here we revisit the common approach of solving the problem of maximizing the log probability of the desired observations by maximizing a variational lower bound, the so-called negative free energy.
We show that, instead, an approach directly maximizing that probability under the agent's policy 
is better suited to, and provides a better solution for, the original homeostatic control problem. 
This is done using hidden Markov model (HMM) control by allowing the policy to act over hidden states or noisy versions thereof while trying to maximize the probability of repeatedly having the desired observations. 
HMM control largely improves performance over the variational, or free energy, approach. 
We also show that the optimal policy is strictly deterministic,
while the variational approach leads to a stochastic policy approximation.
We finally provide a homeostatic reinterpretation of the maximum occupancy principle —a principle proposing that agents ought to maximize the occupancy of action-state path space— 
by defining homeostatic states as any states that do not immediately entail the termination or death of the agent.   
\end{abstract}


\section{Introduction}

Animals strive to maintain their homeostatic levels so that they can survive and act in the world.
Being in a precarious state, an agent needs to regulate internal variables to preserve the integrity of its body. Thus, agents should counteract perturbations pushing them away from desired homeostatic states (e.g., correcting deviations of body temperature toward its ideal value) \cite{keramati2014homeostatic,yoshida2024modeling,petzschner2021computational,yoshida+25-rew-homeostais,yoshida2025unexpected}. 
While some theories of behavior —mostly those based on intrinsic motivation— such as the maximum occupancy principle \cite{moreno2023empowerment,ramirez2024complex} and empowerment \cite{klyubin2005empowerment}, situate motivation signals at the cusp of behavioral causation, other theories, such as drive theory \cite{berridge2004motivation}, place drives and homeostatic needs at the top of the hierarchy \cite{rosado2022drive,juechems2019does}. 
According to this view, behavior naturally emerges from the dynamic regulation of homeostatic needs \cite{berridge2004motivation} (see \cite{moreno2026intrinsic} for a comparison between different behavioral principles).

Several quantitative theories have been put forward to model homeostasis \cite{yoshida+25-rew-homeostais}. A common formalization of homeostasis is the free energy principle \cite{friston2013anatomy,parr2022active}.
This framework incorporates homeostasis by defining desired observations, that is, specific states critical for survival.
The goal of the agent is then to maximize the log probability, or negative surprise, of having the desired observations, so that in most of the time the agent's state is on or close to their ideal homeostatic values. 
In this way, the agent should avoid surprising events that risk homeostatic breakage. 
Solutions to this optimization are found by approximating the log probability objective with a variational lower bound, called negative free energy, which the agent maximizes through updates of its internal states and action policies.

We define homeostatic control as the maximization of the probability of the agent obtaining the desired observations over consecutive time steps, and formulate the problem in terms of hidden Markov model (HMM) control \cite{todorov2009efficient,damiani2024stochastic}.
In HMM control, the agent gets observations from hidden states and acts through a policy that selects actions based on the hidden states or partially observable versions thereof. 
Directly maximizing the probability of the desired observations in this HMM control problem features a simple solution in terms of backpropagated messages.
We compare this solution to the common approach of approximating the maximization of the log probability of the desired observations by a variational lower bound (negative free energy) and then maximizing this lower bound over the policy.  
We find that HMM control improves performance over the variational lower bound approach, in addition to providing a better-matched solution to the original homeostatic control problem. 
We also show that the optimal policy under HMM homeostatic control is strictly deterministic, even when the full problem is effectively an inference problem. 
In contrast, the optimal policy induced by the variational lower bound optimization is strictly stochastic.
We finally provide a homeostatic reinterpretation of the maximum occupancy principle \cite{moreno2023empowerment,ramirez2024complex} by considering homeostatic states as any states that do not entail the immediate termination of the agent's potentiality to act in the world.

\section{The Homeostatic Objective, Variational Approximation, And HMM Control}

We first describe homeostasis as surprise minimization over consecutive time steps, review the free energy approach, and derive the optimal solution under HMM control.

\begin{figure}
\includegraphics[width=14.0cm]{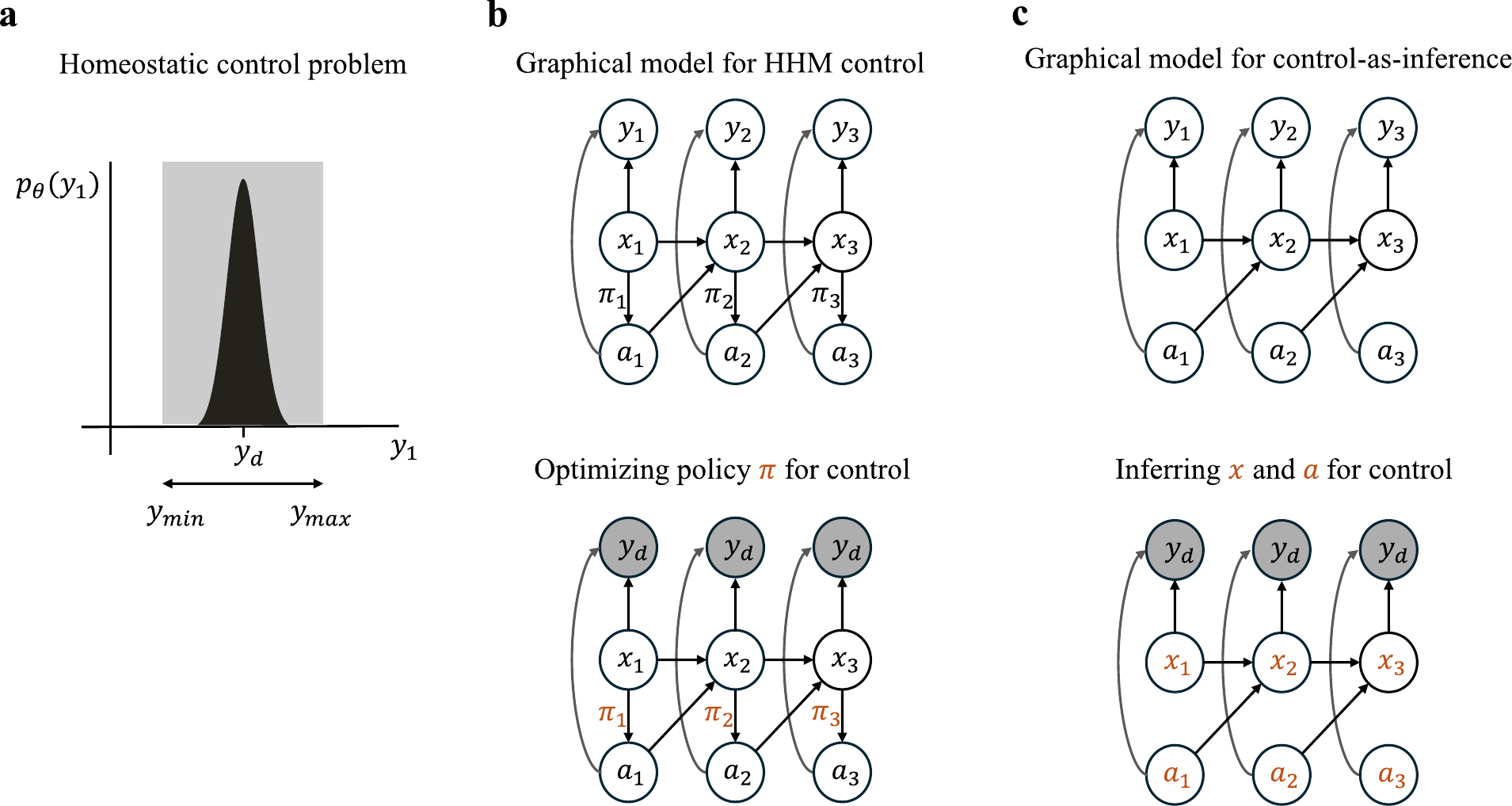}
\centering
\caption{
(\textbf{a}). The homeostatic control problem. The agent must maximize the probability that the observations $y_t$ lie at the desired value $y_d$ for times $t=1,...,T$ (only the probability $p_{\theta}(y_1)$ for $t=1$ shown).
Homeostasis can alternatively be defined by maintaining the observation $y_t$ within some allowed range $(y_{\text{min}},y_{\text{max}})$ (see Sec. \ref{sec:mop}).
(\textbf{b}). Hidden Markov model (HMM) control. The graphical model used to model homeostasis (top panel; with $T=3$ for illustration) consists of an HMM with observations $y_t$, hidden states $x_t$ and actions $a_t$. The graphical model is optimized with respect to the policy $\pi=\pi_{1:T}$ in order to maximize the probability of having the desired observations $y_t=y_d$ over all time steps (bottom).
(\textbf{c}). In control-as-inference approach, the observations $y_t$ are clamped to the desired observations $y_d$, and assuming this to be true, the probability of the sequence of states and actions is inferred (bottom panel). 
This inference process might correspond to states and action sequences that are neither realizable nor likely once the model is run forward in time without clamping the observations to the desired values
\label{fig1}
}
\end{figure}

\subsection{Homeostatic Objective}
\label{sec:homeostatic_objective}

Homeostasis implies the maintenance of internal observations (specific states) within a safe region, such as keeping body temperature and blood glucose levels within healthy ranges \cite{berridge2004motivation,yoshida+25-rew-homeostais}. A common approach is defining a {\em desired} observation value $y_d$ (i.e., the ideal homeostatic state) of a random variable $y_1  \in \mathbb{R}^m$, and an underlying probability distribution over the latter, denoted $p_{\theta}(y_1)$ (for now we only assume a single time step, $t=1$). This probability distribution might depend on the policy of the agent and other factors, and so it is parametrized by a set of values $\theta$ describing these factors. The goal of the homeostatic agent is to maximize the probability $p_{\theta}(y_1=y_d)$ with respect to $\theta$,
\begin{align}
    \theta^* = \arg \max_{\theta} p_{\theta}(y_1=y_d) \;.
    \label{eq:p(y)}
\end{align} 
If the optimized distribution $p_{\theta^{*}}(y_1)$ is sharply peaked at $y_d$ (Fig. \ref{fig1}a), then the agent is guaranteed to lie close to the desired observation $y_d$ with high probability.  
The desired observation $y_d$ can be multidimensional ($m>1$), with each component reflecting different dimensions of homeostatic needs, such as water, sugar, proteins, etc. 

Here, we are concerned with the problem of maintaining homeostasis over time, not just in a single time step. We therefore define a set of random variables $y_t$ at discrete times $t=1,...,T$ with finite horizon $T$, each for simplicity having the same desired value $y_d$. 
We adopt a discrete time Markov Decision Process (MDP) where the agent is in a discrete state $x_t \in X$ at time $t$, chooses a discrete action $a_t \in A$ at that time with probability $\pi_t(a_t|x_t)$ (which defines the agent's {\em policy}), and receives the observation $y_t$ with conditional probability $p(y_t|x_t,a_t)$.
Upon performing action $a_t$ in state $x_t$, the world transitions to state $x_{t+1}$ at the next time step with probability $p(x_{t+1}|x_t,a_t)$. 
Therefore, the observations $y = y_{1:T}$ and states $x = x_{1:T+1}$ form an HMM (Fig. \ref{fig1}b). 
Given this generative model, the joint probability of having states $x$, observations $y$ and generating actions $a=a_{1:T}$ when using the time-dependent policy $\pi=\pi_{1:T}$ is
\begin{align}
    p_{\pi}(y,x,a) = p(x_1) \prod_{t=1}^{T} \pi_t(a_t|x_t) p(y_{t}|x_{t},a_t) p(x_{t+1}|x_t,a_t) \;,
    \label{eq:p_joint}
\end{align}
where $p(x_1)$ is the distribution over the initial state.

To achieve the homeostatic goal in this sequential decision process, the agent should maximize the probability of having the desired observation over the sequence of the $T$ discrete times steps, 
namely, 
\begin{align}
    \pi^* = \arg \max_{\pi} p_{\pi}(y = y_d) \;,
    \label{eq:p(y,y,..,y)}
\end{align}
where the notation $y=y_d$ means that $y_t=y_d$ for all $t$, the optimization now is explicitly performed over the policy $\pi$ (the only quantity that the agent can control), and the marginal probability of having the desired observations is expressed as
\begin{align}
    p_{\pi}(y = y_d) = \sum_{x,a} p_{\pi}(y=y_d,x,a) \;.
    \label{eq:p_y_marginal}
\end{align}
The objective in Eq. \ref{eq:p(y,y,..,y)} implies that choosing a policy that gives a very low probability of having $y_d$ at just a single time step will be strongly discouraged, capturing the idea that for survival the agent needs to maintain homeostasis throughout its lifespan. 

It is important to realize that the optimization problem in Eq. \ref{eq:p(y,y,..,y)} is both a control and an inference problem. First, it is a control problem because the random variables $y_t$ are clamped to the desired value $y_d$, and the agent acts to maximize the probability of having these desired observations by finding a suitable policy $\pi$. In this view, the random variables at test time do not need to visit exactly the desired observations, but the probability of that event should have the highest possible value.
Second, it is also an inference problem because $y_t=y_d$ can be taken as observed data, and then Eq. \ref{eq:p(y,y,..,y)} can be simply understood as a maximum likelihood estimation problem where the "parameters" form arbitrary non-parametric policies $\pi$. In this view, the random variables have been observed, and one wants to find the model parameters (i.e., the policy) maximizing the probability of those observations.
The connection between inference and control problems has been noted several times \cite{toussaint2006probabilistic,todorov2008general,kappen2012optimal,levine2018reinforcement}.

An important question is whether the objective in Eq. \ref{eq:p(y,y,..,y)} can be mapped to the standard approach of MDPs with an additive reward function $r(y_t,a_t)$. 
This mapping, however, is generally not possible, because our objective is neither additive nor perfectly multiplicative. 
Indeed, the probability in Eq. \ref{eq:p(y,y,..,y)} cannot be expressed as a product of marginal probabilities, that is, $p_{\pi}(y = y_d) \neq \prod_t p(y_t=y_d)$, due to the hidden structure of $x$.
This precludes reformulating the objective simply as a sum of log probabilities, $\sum_t \log p(y_t=y_d)$ or expectations over $\sum_t \log p(y_t=y_d|x_t,a_t)$, which would have been handy for a standard reward-based MDP formalization of the problem. 
The imperfect multiplicative nature of the homeostatic objective precludes using the additive formulation based on reward functions considered in standard MDPs, and therefore a Bellman equation associated with our problem is not directly available. 
In Sec. \ref{sec:hmm_control}, we will see that the optimal solution of the objective in Eq. \ref{eq:p(y,y,..,y)} leads to a multiplicative message passing scheme, while reward-based MDPs lead to the well-known additive message passing scheme due to the Bellman recursion. 
 
Although the homeostatic objective in Eq. \ref{eq:p(y,y,..,y)} cannot be cast as a reward-based MDP, the event of obtaining the desired observation $y_t$ can be reinterpreted as a reward, rather than a homeostatic desired observation value. In other words, the condition of having the desired observation $y_t=y_d$ can represent the acquisition of a high reward at time $t$. 
This is clearer if $y_t$ is a binary rather than continuous variable, where $y_t=0$ ($y_t=1$) is interpreted as a low (high) reward.
In this reward-based view, the maximization of the objective in Eq. \ref{eq:p(y,y,..,y)} would imply that the agent aims to maximize the probability of observing a sequence of high rewards, $y=y_d=1$.
Treating the desired observation as a homeostatic target or a high reward leads to the same formulation under Eq. \ref{eq:p(y,y,..,y)}, but we stress again that the reward maximizer constructed in this way is not equivalent to reward-based MDPs with an additive reward-function maximization objective.

Finally, extending our results to continuous action and state spaces is straightforward by just replacing sums over actions and states by integrals. 
Generalizing our results to partial observability of the policy over $x$ is simple, and it is omitted here to avoid unnecessary complications and lengthy equations below.
Our formulation in Eq. \ref{eq:p_joint} follows the standard treatment of partial observability in the active inference literature \cite{isomura2022canonical}.
Extending all the theory to time-dependent state transition matrices, observation models and desired observation values can readily be done by inserting time indexes.

\subsection{Variational Inference and Free Energy}
\label{sec:variational_inference}

As explained before, the objective defined in Eq. \ref{eq:p(y,y,..,y)} can be cast as an inference problem, consisting of maximizing the likelihood of the observation $y_d$ over the policy $\pi = \pi_{1:T}$, where the policy is a parametric (or non-parametric) probability distribution with time-dependent parameters $\theta_t$.
Thus, methods to solve the above optimization problem exploit common approaches in graphical models and probabilistic inference, such as variational inference \cite{bishop2006pattern,parr2022active}. 

Variational inference becomes a convenient approach when optimizing parameters in an inference problem involves marginalization over hidden states in large spaces \cite{bishop2006pattern}.
Marginalization is intractable in very large problems, and variational approaches often allow analytically tractable marginalization, however, at the cost of providing a suboptimal solution. 
As computing the probability of obtaining the desired observations in Eq. \ref{eq:p_y_marginal} involves such a marginalization, variational presents itself as a good approach in our homeostatic problem as well
(later in Sec. \ref{sec:hmm_control} we exploit the underlying sequential nature of the optimization of the marginal distribution, providing a separate route to its solution). 

Variational inference approximates the optimal policy by minimizing the divergence between a target and variational distribution. This renders the problem in Eqs. \ref{eq:p(y,y,..,y)}-\ref{eq:p_y_marginal} into a different but related one \cite{levine2018reinforcement}, an important point that will be discussed later. 
We start by defining a probabilistic model with probability distribution over observations, states and actions
\begin{align}
    p(y,x,a) \propto p(x_1) \prod_{t=1}^{T} p(y_{t}|x_{t},a_t)^\alpha \; p(x_{t+1}|x_t,a_t)   \;,
    \label{eq:p_joint_approx}
\end{align}
where we note that the assumed policy is uniform over the actions and fixed, so we omit the conditioning on the policy $\pi$. 
The parameter $\alpha$ controls how strongly the the states $x_t$ should be pushed to produce the desired observation values $y_t=y_d$.
A natural choice is $\alpha=1$ because it is when the graphical model resembles most the original model, Eq. \ref{eq:p_joint}.
In the limit $\alpha \rightarrow \infty$, the model only gives non-zero probability to those $x_t$ with the maximum probability of producing the desired observation. 
Therefore, when conditioning on the desired observations, $y=y_d$,
this probabilistic graphical model (Fig. \ref{fig1}c) gives high probability weight to states and actions producing the desired observations. 
Methods that infer the policy using an underlying graphical model are often referred to as {\em control-as-inference}. 

A key idea in control-as-inference is that the posterior over actions given current state and desired observations obtained from the probabilistic model in Eq. \ref{eq:p_joint_approx}, and denoted $\gamma_t(a_t|x_t,y=y_d)$, can form a good approximation to the optimal policy $\pi_t(a_t|x_t)$ in the homeostatic control problem defined by Eq. \ref{eq:p(y,y,..,y)}. 
This action posterior can be computed from the joint probability of states and actions given the desired observations, denoted $\gamma(x,a|y=y_d)$, using Bayes rule in Eq. \ref{eq:p_joint_approx}.
The joint probability $\gamma(x,a|y=y_d)$ dictates both the actions and states that need to be visited when conditioning the observations on the desired ones, that is, what the agent should do and visit conditioned on having the desired observation $y_t=y_d$ at every time.
Although reasonable, this probability distribution assumes excessive control over the world dynamics, as it is the joint over $x,a$ which is obtained, but the agent actually only has control over the actions $a$ and not exactly over the states $x$ that will be visited —this is one place where one can see that inferring the hidden states $x$ is not quite the same as actually controlling them, making inference and control not identical 
\footnote{This situation can be illustrated with the following example of an agent that wants to be rich in the future: in order to maximize the probability of getting rich, the agent infers the best states and actions that should be visited and taken; however, this inference process does not guarantee that the probability of getting rich will be high, and, even more dramatically, this wishful thinking can lead to courses of action ending with a high probability of becoming extremely poor.}
\footnote{This problem may be exacerbated in mean-field approximations of variational inference in active inference, as then the transition dynamics between states is left unconstrained.}.
The implication is that the actual policy $\pi(a_t|x_t,y=y_d)$, obtained using Bayes rule from $\gamma(x,a|y=y_d)$, will produce risky behavior, as it assumes that it is possible to control the state $x$ beyond what the world dynamics affords.
Specifically, using Bayes rule we could obtain $p(x_{t+1}|x_t,a_t,y=y_d)$, but this transition probability might not be the same as the actual transition probability $p(x_{t+1}|x_t,a_t)$ \cite{levine2018reinforcement} (see Sec. \ref{sec:comparing} for a deeper analysis). 

To constrain the agent to be able to select only actions, and not directly states, we approximate the ideal $\gamma(x,a|y=y_d)$ from the probabilistic model defined in Eq. \ref{eq:p_joint_approx} by another distribution $q_{\pi}(x,a)$ (the so-called variational approximation)
\begin{align}
    q_{\pi}(x,a) = p(x_1) \prod_{t=1}^{T} \pi_t(a_t|x_t) p(x_{t+1}|x_t,a_t) \;,
    \label{eq:p_joint_var_approx}
\end{align}
which follows the same dynamical structure as the problem and that allows optimizing the policy $\pi$ —and only the policy.
The variational approximation $q_{\pi}(x,a)$ does not contain the observation probabilities because it is the joint $\gamma(x,a|y=y_d)$ conditioned on the desired observations what we want to approximate; we do not aim at approximating the full joint $p(y,x,a)$.
Note that the variational approximation permits only the control over actions, and thus only partial control over state transitions through the world dynamics $p(x_{t+1}|x_t,a_t)$. Therefore, $q_{\pi}(x,a)$ is a realizable joint over states and actions given the assumptions of what the agent can do in the world, in contrast to what $\gamma(x,a|y=y_d)$ implies.  

In summary, the variational inference approach chooses the policy $\pi$ so as to minimize the difference between the variational approximation $q_{\pi}(x,a)$ and the target joint probability distribution $\gamma(x,a|y=y_d)$ from the probabilistic model in Eq. \ref{eq:p_joint_approx}. This approach implements the idea that the variational approximation respects the partial controllability of the world dynamics and that the target joint distribution represents the ideal control and therefore we would like to be as close to it as possible. 
In mathematical form, we optimize $\pi$ so as to minimize the Kullback-Leibler divergence
\begin{align}
    & D_{\text{KL}}(q_{\pi}||\gamma) = \sum_{x,a} q_{\pi}(x,a) \log \frac{q_{\pi}(x,a)}{\gamma(x,a|y=y_d)} \;,
    \label{eq:KL}
    \\
    &  \pi^* = \arg \min_{\pi} D_{\text{KL}}(q_{\pi}||\gamma) 
    = \arg \min_{\pi} E_{(x,a) \sim q_{\pi}} \log \frac{q_{\pi}(x,a)}{p(y=y_d,x,a)}\;,
    \label{eq:min_KL}
\end{align}
where in the right hand side of the last line we have used that $\gamma(x,a|y=y_d)$ and the joint $p(y=y_d,x,a)$ in Eq. \ref{eq:p_joint_approx} differ by the constant $p(y=y_d)$, and $E_{(x,a) \sim q_{\pi}}$ means expectation with respect to $q_{\pi}(x,a)$.

Therefore, using Eqs. \ref{eq:p_joint_approx}, \ref{eq:p_joint_var_approx}, \ref{eq:min_KL} and changing signs, the optimal policy can be calculated as 
\begin{align}
    \pi^* & = \arg \max_{\pi} E_{x,a \sim q_{\pi}} \sum_t \left( -\log \pi_t(a_t|x_t) + \alpha \log p(y_t=y_d|x_t,a_t) \right)
    \nonumber
    \\
    & =
    \arg \max_{\pi} \sum_{x,a} q_{\pi}(x,a) \sum_t \left( -\log \pi_t(a_t|x_t) + \alpha \log p(y_t=y_d|x_t,a_t) \right) \;.
    \label{eq:KL_expanded}
\end{align}
This is the standard objective for a reward-based MDP with reward function $r(x_t,a_t) = \alpha \log p(y_t=y_d|x_t,a_t) -\log \pi_t(a_t|x_t)$, where the first term is an extrinsic reward that is high when close to the desired observation, and the second term is an action entropy regularization term \cite{haarnoja2018soft}.
The extrinsic reward term favors encountering states $x_t$ for which the probability of having the observation $y_d$ is high. 
This effect is exacerbated by increasing $\alpha$, so that the action entropy regularization stops having any role and the problem converges to that of pure reward maximization with reward proportional to $\log p(y_{t}=y_d|x_t,a_t)$.
This implements an approximate homeostatic mechanism of staying close to the desired observations at every time.

The optimal policy in this MDP can be obtained using the associated Bellman equation as \cite{todorov2009efficient}
\begin{align}
    \pi^*_t(a_t|x_t) & \propto 
    p(y_{t}=y_d|x_t,a_t)^\alpha
    \exp \left( \sum_{x_{t+1}} p(x_{t+1}|x_t,a_t) V^*_{t+1}(x_{t+1}) \right) 
    \label{eq:optimal_policy}
    \\
     V^*_t(x_t) & = \log \left[ \sum_{a_t} p(y_{t}=y_d|x_t,a_t)^\alpha 
     \exp \left( \sum_{x_{t+1}} p(x_{t+1}|x_t,a_t) V^*_{t+1}(x_{t+1}) \right) \right]  \;,
    \label{eq:optimalvalue}
\end{align}
where $V^*_t(x_t)$ is the optimal value function when the agent is in state $x_t$ at time $t$, computed recursively backward in time with the boundary condition $V^*_{T+1}(x_{T+1})=0$ for all $x_{T+1}$.

We recall that the original objective was maximizing the probability of obtaining the observation $y_t=y_d$ for all $t$, Eq. \ref{eq:p(y,y,..,y)}. Here, we find that this objective can be approximated by a MDP with reward equal to $r(x_t,a_t) \simeq  \alpha \log p(y_t=y_d|x_t,a_t)$. These two problems are not quite the same, because $\log p_{\pi}(y=y_d)$ is not equal to $E_{x,a \sim q_{\pi}} \sum_t \log p(y_t=y_d|x_t,a_t)$, as we have argued in Sec. \ref{sec:homeostatic_objective}.
However, we can see that optimizing $\pi$ under the MDP problem in Eq. \ref{eq:KL_expanded} provides a lower bound to the log probability of the observations in Eq. \ref{eq:p(y,y,..,y)}, the original problem, since (expanding all standard steps for clarity)
\begin{align}
    \log p(y=y_d) & = \log \sum_{x,a} p(y=y_d,x,a) = \log \sum_{x,a} p(y=y_d,x,a) \frac{q_{\pi}(x,a)}{q_{\pi}(x,a)}
    \nonumber
    \\
    & = \log \sum_{x,a} q_{\pi}(x,a) \frac{p(y=y_d,x,a)}{q_{\pi}(x,a)}
    \nonumber
    \\
    & \geq  E_{x,a \sim q_{\pi}} \log \frac{p(y=y_d,x,a)}{q_{\pi}(x,a)}
    \nonumber
    \\
    & = \log p(y=y_d) - D_{\text{KL}}(q_{\pi}||\gamma) \; ,
    \label{eq:p(y)_bound}
\end{align}
where Jensen inequality has been used in the third line. 
Two important well-known facts can be extracted from here \cite{buckley2017free}.
First, the expectation $- F = E_{x,a \sim q_{\pi}} \log ( p(y=y_d,x,a) / q_{\pi}(x,a) )$ (see third line) forms a lower bound on the log probability of the desired observation. 
This bound is called the evidence lower bound (ELBO) in the machine learning literature \cite{Blei_2017}, and the negative free energy in the active inference theoretical literature \cite{parr2019generalised}.
Second, the divergence $D_{\text{KL}}(q_{\pi}||\gamma)$ must be non-negative, $D_{\text{KL}}(q_{\pi}||\gamma) \geq 0$, and it is zero only when $q_{\pi}(x,a)$ matches $\gamma(x,a|y=y_d)$. 
We have argued above that matching $q_{\pi}$ to $\gamma$ is not achievable, because it would involve having unrealistic control over the states. 
Nevertheless, optimizing $\pi$ involves approximating $q_{\pi}$ as closely as possible to the ideal $\gamma$ while at the same time respecting the partial controllability of the state transitions through the agent's actions. 

In conclusion, several key distinctions emerge. First, the optimal policy obtained under Eq. \ref{eq:KL_expanded} is stochastic (as shown in Eq. \ref{eq:optimal_policy}) due to the entropy regularizer in the objective \cite{todorov2009efficient,moreno2023empowerment}. Second, while our formulation aligns with the standard variational control-as-inference framework and its associated graphical model \cite{levine2018reinforcement}, it differs from mainstream active inference literature, where policies are typically modeled as open-loop action sequences \cite{parr2019generalised}. Here, we instead adopt the standard closed-loop policy framework of MDPs. Finally, incorporating explicit epistemic value alongside the homeostatic objective does not alter our core theoretical conclusions.


\subsection{HMM Control}
\label{sec:hmm_control}

Variational inference is one possible route to solve the homeostatic control problem in Eq. \ref{eq:p(y,y,..,y)}. 
However, this approach may be too generic to exploit the full underlying structure of the problem. 
In this section we show that it is indeed possible to obtain an improved solution and policy by exploiting the underlying sequential structure of the HMM and the flexibility afforded to the agent to control actions at will. 
This leads to a natural derivation of the optimal policy without the need to optimize any variational lower bound or assuming a different probabilistic graphical model from the one defined in the original problem. 

To obtain the optimal policy for the homeostatic control problem, we first recall that the objective in Eq. \ref{eq:p(y,y,..,y)} can be written using Eq. \ref{eq:p_joint} as 
\begin{align}
    \pi^* & = \arg \max_{\pi} p_{\pi}(y=y_d) 
    \nonumber
    \\
    & = \arg \max_{\pi} \sum_{x,a} p(x_1) \prod_{t=1}^{T} \pi_t(a_t|x_t) p(y_{t}=y_d|x_{t},a_t) p(x_{t+1}|x_t,a_t)   \;.
    \label{eq:p_pi_direct_optimization}
\end{align}
This objective is linear in each policy $\pi_t$ at time $t$, and because policies form simplexes, they must attain their maximum at their boundaries.
This in turn implies that there must be an optimal deterministic policy. 
Therefore, a first important conclusion is that although the optimal policy in the variational inference approach is strictly stochastic (see Eq. \ref{eq:optimal_policy}), the actual optimal policy is deterministic. 
Thus, the policy obtained in Eq. \ref{eq:optimal_policy} cannot be optimal under the original homeostatic control problem. 
This result can be understood simply because variational inference assumes a "rigid" graphical model with a fixed policy (uniform in the case considered, Fig. \ref{fig1}c, but it can also be any non-uniform policy, leading to similar results \cite{todorov2009efficient}) and finds an "optimal" policy by approximating the posterior over states and actions in this model. In contrast, the original homeostatic problem has a "flexible" graphical model where the policy can be optimized (Fig. \ref{fig1}b). 

The optimal policy $\pi$ can be directly obtained from Eq. \ref{eq:p_pi_direct_optimization} backward in time, because $\pi_t$ only affects future states, and not current or past ones. 
Starting at time $t=T$, the optimal action is
\begin{align}
    a^*_{T}(x_{T}) = \arg \max_{a_T} p(y_T=y_d|x_T,a_T) \;.
    \nonumber
\end{align}
We define the message at time $T$ as
\begin{align}
    \beta_{T}(x_{T}) = p(y_T=y_d|x_T,a^*_T(x_T)) \;.
    \nonumber
\end{align}
Solving backward in time, the optimal actions and messages are recursively computed as
\begin{align}
    a^*_{t}(x_{t}) & = \arg \max_{a_t}  p(y_{t}=y_d|x_t,a_t) \sum_{x_{t+1}} p(x_{t+1}|x_{t},a_t) \beta_{t+1}(x_{t+1})
    \label{eq:policy_beta}
    \\
    \beta_{t}(x_{t}) &= p(y_{t}=y_d|x_t,a^*_{t}(x_{t})) \sum_{x_{t+1}} p(x_{t+1}|x_{t},a^*_{t}(x_{t})) \beta_{t+1}(x_{t+1}) \;.
     \label{eq:message_beta}
\end{align}
This recursion defines a deterministic policy $\pi_t$ for every time step $t$.

The backward messages $\beta$ play the role of the value function in the Bellman equation, but the recursion takes the form of a multiplication rather than the standard addition in the Bellman backup. 
Indeed, the messages are probabilities. 
Specifically, $\beta_t(x_t) = p(y_{t:T}=y_d|x_t)$ is the probability of observing the desired values from $t$ onward when following the optimal policy starting at $t$.
The optimal policy at time $t$ in Eq. \ref{eq:policy_beta} chooses the actions that maximize the probability of obtaining the desired observation from $t$ up to $T$.
Backward message schemes with multiplicative nature similar to ours also arise as control-as-inference in the expectation step in an EM-like approach for a fully observable MDP problem with additive rewards \cite{toussaint2006probabilistic}, which differ from our setting. 
They also arise in control-as-inference \cite{levine2018reinforcement} without the maximum over policies operator because a fixed graphical model is considered (see Fig. \ref{fig1}c). 

The deterministic nature of the solution of the HMM control problem defined by Eqs. \ref{eq:p_pi_direct_optimization}-\ref{eq:message_beta} remains unaltered if the policy can only act on partial observability of the hidden state $x$, rather than directly over it. 
In this case, the policy is constrained to be conditioned on the observation $z_t = g(x_t)$, that is, $\pi_t(a_t|z_t)$, where $g$ is a many-to-one function, so that many (at least more than one) hidden states lead to the same observation for the policy. 
The sums in Eqs. \ref{eq:p_pi_direct_optimization}-\ref{eq:message_beta} become slightly more cumbersome, but the linearity of the objective \ref{eq:p_pi_direct_optimization} in the policies $\pi_t$ persists, and hence the optimality of a deterministic policy remains.

\subsection{Homeostatic Control Through The Maximum Occupancy Principle}
\label{sec:mop}

The objective of maximizing the probability of desired observations for homeostatic control, as described in Eq. \ref{eq:p(y,y,..,y)}, is appealing. 
However, it suffers from the limitation that one needs to define the desired observation values $y_d$ themselves. 
In practice, a homeostatic control system might be better conceptualized as allowing for a sometimes wide range of valid values, not a single strict value. 
However, extending the formalism to include ranges would require using additional partial marginals over the observation probabilities of Eq. \ref{eq:p(y,y,..,y)}, which would complicate the problem solution. 

To avoid additional marginalization over ranges of allowed homeostatic values, we now focus on the boundaries of the homeostatic range, and treat these boundaries as terminal observations, meaning observations that imply the termination, or death, of the agent. 
For instance, in the case of body temperature, we may define a healthy range, and also the boundaries that do not allow survival, i.e., too low or too high temperature (see Fig. \ref{fig1}a). 
Therefore, the agent must act to avoid those terminal observations, but it is left without a task if the temperature is within the allowed range.
In this sense, the agent is indifferent regarding what value the observation takes within this range. 

To provide the agent with a "task", we propose that the agent maximizes the entropy of the sequences of actions that it generates and the observations it obtains while avoiding the terminal observations.
This objective does not imply that every action and observation is visited with the same probability, as getting close to a terminal observation should be disfavored just due to the proximity to death.
The objective of maximum action-observation path occupancy is known as the Maximum Occupancy Principle (MOP) \cite{moreno2023empowerment,ramirez2024complex}, and it has been argued that this intrinsic motivation to generate path complexity better describes complex behavior in nature than purely reactive homeostasis \cite{moreno2026intrinsic}. 
 
Formally, the agent's objective is to optimize the policy $\pi$ so as to maximize the sum of action and observation entropies,
\begin{align}
    V_{\pi} = E_{(x,a) \sim q_{\pi}, y \sim p(y|x,a)} \sum_{t=1}^T \left( -\log \pi_t(a_t|x_t) - \beta \log p(y_t|x_t,a_t) \right) \;,
    \label{eq:V_mop}
\end{align}
where the expectation is over $q_{\pi}$, defined in Eq. \ref{eq:p_joint_var_approx}, and over observations $y_t \sim p(y_t|x_t,a_t)$, and $\beta$ is a parameter controlling the relative importance of the observation vs. action entropies ($\beta \ge 0$ typically, but it can also be chosen negative, $\beta < 0$, to provide an approximation to Empowerment \cite{jung2011empowerment}, which prefers controllable state transitions over stochastic ones). 
To capture the homeostatic need of avoiding the terminal observation, the episode terminates at an earlier time $t<T$ whenever $y_t$ lies outside the allowed range, $y_t \notin y_{\text{safe}} =  (y_{\text{min}},y_{\text{max}})$.
Note that $y_t$ is not clamped at any desired value, but rather it is considered a random variable generating the reward $-\beta \log p(y_t|x_t,a_t)$. 

The new objective in Eq. \ref{eq:V_mop} is superficially similar to Eq. \ref{eq:KL_expanded}, but with the important difference that now the observation $y_t$ is not fixed at a desired value.
Therefore, the second term $-\log p(y_t|x_t,a_t)$ becomes a reward function that encourages observation entropy, that is, discourages observations that have high probability, quite the opposite to classical homeostasis.
This is analogous to the entropy of the actions, which is promoted by the first term $-\log \pi(a_t|x_t)$.
Altogether, the agent seeks to maximize action-observation path entropy over the course of its lifespan (providing complexity), at the same time staying within the safe observation range (providing structure through its embodiment).

The optimal policy for the MOP objective in Eq. \ref{eq:V_mop} can be computed recursively backward in time \cite{todorov2009efficient,ramirez2024complex} as
\begin{align}
    \pi^*_t(a_t|x_t) & \propto 
    \exp  \left( \beta H(Y|x_t,a_t) + p_{\text{safe}}(x_t,a_t) \sum_{x_{t+1}} p(x_{t+1}|x_t,a_t)  V^*_{t+1}(x_{t+1})  \right) 
    \label{eq:optimal_policy_mop}
    \\
     V^*_t(x_t) & = \log \left[ \sum_{a_t} \exp \left( \beta H(Y|x_t,a_t) +  p_{\text{safe}}(x_t,a_t) 
     \sum_{x_{t+1}} p(x_{t+1}|x_t,a_t) V^*_{t+1}(x_{t+1}) \right) \right] \;,
    \label{eq:optimalvalue_mop}
\end{align}
where $H(Y|x_t,a_t) = - \sum_y p(y|x_t,a_t) \log p(y|x_t,a_t) $ is the observation entropy, $p_{\text{safe}}(x_t,a_t) = \int_{y_{\text{min}}}^{y_{\text{max}}} dy \; p(y|x_t,a_t) $ is the probability of having a safe observation at time step $t$ when at state $x_t$ and performing action $a_t$, and $V^*_t(x_t)$ is the optimal value function from state $x_t$ at time $t$, computed with boundary condition $V^*_{T+1}(x_{T+1})=0$ for all $x_{T+1}$. 
Note that $p_{\text{safe}}(x_t,a_t)$ acts as a continuation probability.
The boundary condition on terminal observations gives MOP a rich structure in the optimal policy, capturing the embodiment of the agent and precluding the policy from being purely random.

MOP presupposes that all states within the wide range of homeostatically controllable values are a priori equally desirable. This does not mean that MOP is equivalent to making the probability of observations flat over a broad range of values in the homeostatic control problem defined by Eq. \ref{eq:p(y,y,..,y)}. Doing this would produce an optimal policy that is deterministic, as shown in the previous section, while the optimal policy in MOP (Eq. \ref{eq:optimal_policy_mop}) is inherently stochastic.
Therefore, the objective in MOP notably differs from classical homeostasis but captures the idea that staying within a safe range is a mandatory boundary condition.   

\section{Comparing HMM control, control-as-inference and variational inference}
\label{sec:comparing}

Next, we more deeply compare HMM control, control-as-inference and variational inference for control in a simple setting and example. We first show that if the horizon is $T=1$ and $\alpha \rightarrow \infty$, then the optimal policy is identical for the three approaches. The first case where differences can be observed is for horizon $T=2$. For $\alpha < \infty$, HMM control has an optimal deterministic policy, while both inference approaches are stochastic, and thus the approaches are different in nature. Therefore, we restrict ourselves here to the case $\alpha \rightarrow \infty$, where we will show that there are optimal deterministic policies for the three approaches. This analysis provides further intuitions on why control-as-inference induces risky behavior.

Starting with the case $T=1$, HMM control maximizes 
\begin{align}
     p_{\pi_1}(y_1=y_d) = \sum_{x_1,a_1} p(x_1) \pi_1(a_1|x_1) p(y_1=y_d|x_1,a_1)
     \;,
    \label{eq:HMM_objective_T=1}
\end{align}
with respect to the policy $\pi_1(a_1|x_1)$ (see Eq. \ref{eq:p_y_marginal}), and has an optimal deterministic policy 
\begin{align}
     a^*_{1,\text{HMM}}(x_1) = \arg \max_{a_1} p(y_1=y_d|x_1,a_1)
     \;.
    \label{eq:HMM_policy_T=1}
\end{align}
Control-as-inference defines the joint distribution
\begin{align}
     p(y_1=y_d, x_1, a_1) = p(x_1) p(y_1=y_d|x_1,a_1)^{\alpha}
     \;
    \label{eq:contro-as-inf_objective_T=1}
\end{align}
(see Eq. \ref{eq:p_joint_approx}).
Using Bayes rule, the inferred policy is stochastic and takes the form
\begin{align}
     \gamma(a_1|x_1) \propto  p(y_1=y_d|x_1,a_1)^{\alpha}
     \;.
    \label{eq:control_as_inf_policy_T=1_alpha}
\end{align}
In the limit $\alpha \rightarrow \infty$, the optimal policy becomes deterministic and identical to the HMM control policy, Eq. \ref{eq:HMM_policy_T=1}.
Under the variational inference approach, the expected reward for $\alpha \rightarrow \infty$ is proportional to  
\begin{align}
    R_{\pi_1} = \sum_{x_1,a_1} p(x_1) \pi_1(a_1|x_1) \log p(y_1=y_d|x_1,a_1)
     \;
    \label{eq:var_inf_objective_T=1}
\end{align}
(see Eq. \ref{eq:KL_expanded}), which again has the same optimal policy as HMM control.
Therefore, the three approaches have the same optimal deterministic policy for $T=1$.

For $T=2$, HMM control now has the objective 
\begin{align}
     & p_{\pi_1,\pi_2}(y_1=y_d, y_2=y_d) 
     \nonumber
     \\
       & \;\;\;\;\;\; = \sum_{x_1,a_1,x_2,a_2} p(x_1) \pi_1(a_1|x_1) p(y_1=y_d|x_1,a_1) p(x_2|x_1,a_1) \pi_2(a_2|x_2) p(y_2=y_d|x_2,a_2) 
     \;,
    \label{eq:HMM_objective_T=2}
\end{align}
with the optimal deterministic policy at time $t=1$
\begin{align}
     a^*_{1,\text{HMM}}(x_1) = \arg \max_{a_1} p(y_1=y_d|x_1,a_1) \sum_{x_2} p(x_2|x_1,a_1) p(y_2=y_d|x_2,a^*_2(x_2))
     \;,
    \label{eq:HMM_policy_T=2}
\end{align}
where we have used the fact that at time $t=T=2$ the optimal deterministic policy, denoted $a^*_2(x_2)$, is given by Eq. \ref{eq:HMM_policy_T=1}.

Control-as-inference defines the joint distribution
\begin{align}
     &p(y_1=y_d, y_2=y_d, x_1, x_2, a_1, a^*_2(x_2)) 
     \nonumber
     \\
     & \;\;\;\;\;\; = p(x_1) p(y_1=y_d|x_1,a_1)^{\alpha} \; p(x_2|x_1,a_1) p(y_2=y_d|x_2,a^*_2(x_2))^{\alpha}
     \;,
    \label{eq:control_as_inf_objective_T=2}
\end{align}
where again we have used that the optimal policy at $t=T=2$ is identical to that of HMM control when $\alpha \rightarrow \infty$ .
Using Bayes rule, the inferred policy becomes
\begin{align}
     \gamma(a_1|x_1) \propto  p(y_1=y_d|x_1,a_1)^{\alpha}
            \sum_{x_2} p(x_2|x_1,a_1) p(y_2=y_d|x_2,a^*_2(x_2))^{\alpha}
     \;.
    \label{eq:control_as_inf_policy_T=2_alpha}
\end{align}
Now the limit $\alpha \rightarrow \infty$ is trickier than for $T=1$.
Among the successor states $x_2$ from $x_1$, the sum will only give substantial weight to the $x_2$ value that maximizes $p(y_2=y_d|x_2,a^*_2(x_2))$ (denoted $x^*_2$), because $\alpha$ grows to infinity, even if $p(x^*_2|x_1,a_1)$ has very small (but nonzero) probability. 
Assuming that $x^*_2$ is unique, in the limit the optimal policy becomes 
\begin{align}
     a^*_{1,\text{CaI}}(x_1) 
     = \arg \max_{a_1}  p(y_1=y_d|x_1,a_1) \; p(y_2=y_d|x^*_2(x_1,a_1),a^*_2(x^*_2))
     \;,
    \label{eq:control_as_inf_policy_T=2}
\end{align}
with the transition probability $p(x_2^*|x_1,a_1)$ only determining the result among ties. 
Although this policy is deterministic, its form is different from HMM control, Eq. \ref{eq:HMM_policy_T=2}, and therefore HMM control and control-as-inference are not in general identical for $T=2$ even in the large-$\alpha$ limit. Therefore, it should be possible to find examples in which the optimal actions taken by each policy differ.
The next example shows this, in addition providing insights into the risk-seeking behavioral attitude of control-as-inference. 

Assume only two possible actions, $a_1$ and $a_1'$, and that the probability of transitioning to $x^*_2$ is $>0$ only when choosing $a_1$ (that is, it is not possible to transition to $x^*_2$ if choosing $a_1'$).
Then, the action $a_1$ will be preferred by control-as-inference, and this preference would be independent of how small the transition probability $p(x^*_2|x_1,a_1)$ is (see Eq. \ref{eq:control_as_inf_policy_T=2}, assuming $p(y_1=y_d|x_1,a_1)$ is independent of the action $a_1$).
However, this is not the case for the optimal policy in HMM control.
To see this, in addition to the above first assume that the successor states $x_2$ after performing $a_1$ are such that
$p(y_2=y_d|x_2,a_2^*(x_2))=0$ for all $x_2$ except for $x_2=x_2^*$ (that is, $a_1$ is a risky action).  
Second, assume that although action $a_1'$ cannot lead to $x^*_2$, it leads with probability one to another state $x_2'$ for which $p(y_2=y_d|x_2',a_2^*(x_2')) > 0$ (that is, $a_1'$ is a safe action). 
Then, for small enough $p(x^*_2|x_1,a_1)$, HMM control will prefer action $a_1'$ over $a_1$ because the latter gives negligible probability mass in the sum of Eq. \ref{eq:HMM_policy_T=2}.
Therefore, HMM control is generally very sensitive to the transition probability to a given state, unlike control-as-inference. 
In sum, inferring the state-action trajectories leading to the highest probability of obtaining the desired observations and implementing the associated policy in the world, as control-as-inference does, assumes too much control over the world dynamics. This process corresponds to a form of wishful thinking. 


Finally, for $T=2$, in the variational inference approach the expected reward for $\alpha \rightarrow \infty$ is proportional to  
\begin{align}
    R_{\pi_1} = \sum_{x_1,a_1} p(x_1) \pi_1(a_1|x_1) \left[ \log p(y_1=y_d|x_1,a_1)
       + \sum_{x_2} p(x_2|x_1,a_1) \log p(y_2=y_d|x_2,a^*_2(x_2)) \right]
     \;,
    \label{eq:var_inf_objective_T=2}
\end{align}
with optimal policy 
\begin{align}
    a_{1,\text{VI}}^*(x_1) = \arg \max_{a_1}  \left[ \log p(y_1=y_d|x_1,a_1)
       + \sum_{x_2} p(x_2|x_1,a_1) \log p(y_2=y_d|x_2,a^*_2(x_2)) \right]
     \;.
    \label{eq:var_inf_policy_T=2}
\end{align}
This optimal policy differs from those in HMM control and control-as-inference. 
However, we can take logarithms in the expression for the HMM control policy in Eq. \ref{eq:HMM_policy_T=2} and rewrite it as 
\begin{align}
    a_{1,\text{HMM}}^*(x_1) = \arg \max_{a_1}  \left[ \log p(y_1=y_d|x_1,a_1)
       + \log \left( \sum_{x_2} p(x_2|x_1,a_1) p(y_2=y_d|x_2,a^*_2(x_2)) \right) \right]
     \;.
    \label{eq:HMM_policy_T=2_new}
\end{align}
These last two expressions are very similar, only differing in where the sum enters, outside or inside the logarithm. This is enough to cause the actual optimal actions to diverge between HMM control and variational inference: if there is an optimal action $a_{1,\text{HMM}}^*$ in HMM control for which a transition to a state $x_2$ with $p(y_2=y_d|x_2,a^*_2(x_2))=0$ is possible and no other zero of $p(y_2=y_d|x_2,a_2)$ occurs,
then $a_{1,\text{HMM}}^*$ will never be chosen by the variational inference controller due to the possibility of receiving an $-\infty$ reward. 

It is easy to check that under deterministic world dynamics, HMM control, control-as-inference and variational inference are identical for all $T$.

\section{The advantage of HMM control over variational inference}

\begin{figure}[H]
\includegraphics[width=14.0 cm]{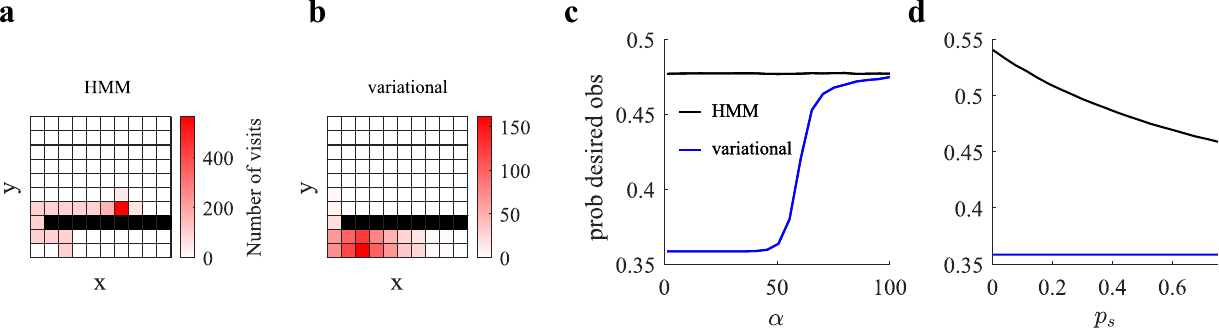}
\centering
\caption{
(\textbf{a, b}). Visitation count for HMM control (\textbf{a}) and variational inference control (\textbf{b}; $\alpha=1$) in a grid world with 5 possible actions (staying, or moving in any cardinal direction one step), except when moving towards the impenetrable walls (black cells). 
HMM control is deterministic, while variational inference control is stochastic. The location with the highest probability of having the desired observation ('best location', reddest cell, at $(8,5)$) is encountered and frequently visited by the HMM agent, but not by the variational inference agent. At the best location $p(y_t=y_d|x_t=x_{best},a_t) = 1$, regardless of time or action, while at other locations $p(y_t=y_d|x_t,a_t) = 0.95$. We used $T=20$, and $100$ trajectories to compute heat maps. At the best location, the action 'staying' results in randomly choosing any of the cardinal directions with total probability $p_s=0.5$.
(\textbf{c, d}). Probability of consecutively obtaining the desired observation, $p(y=y_d)$, as a function of $\alpha$ with $p_s=0.5$ (\textbf{c}) and $p_s$ with $\alpha=1$ (\textbf{d}). Shaded areas, almost the same size of the line widths, correspond to s.e.m. over $10^4$ episodes.
\label{fig2}
}
\end{figure}

The previous sections establish our two main results: (1) Maximizing the probability of the desired observation in Eq. \ref{eq:p(y,y,..,y)} is exactly solved by HMM control, Eqs. \ref{eq:policy_beta}-\ref{eq:message_beta}, while variational inference (equivalently, free energy), Eq. \ref{eq:KL_expanded}, provides a generally suboptimal policy, and (2) HMM control policy is deterministic, Eq. \ref{eq:policy_beta}, while the variational inference policy is strictly stochastic, Eq. \ref{eq:optimal_policy}, when $\alpha < \infty$.
Therefore, regardless of the specific problem considered, substantial differences between the HMM control and variational inference solutions are expected.

Here we illustrate these generally valid conclusions in a simple grid-world example (Fig. \ref{fig2}). The agent receives the desired observation with high probability $0.95$ everywhere in the grid, and receives it with probability $1$ at the best location. The HMM agent follows a deterministic policy toward the best location (Fig. \ref{fig2}a), while the variational agent has a stochastic policy and gets stuck around the initial location and almost never leaves the lower room (Fig. \ref{fig2}b).
The HMM agent attains a higher probability of receiving the desired observation than the variational agent (Fig. \ref{fig2}c,d). 
When $\alpha$ is very large, the variational agent's policy becomes nearly deterministic and approaches the HMM optimal policy (Fig. \ref{fig2}c). This result is coincidental, as generally the HMM and variational optimal policies differ in the limit $\alpha \rightarrow \infty$, although both become deterministic (see Sec. \ref{sec:comparing}). 
Varying the action noise probability $p_s$ at the best location affects the HMM agent but not the variational one (with $\alpha=1$).

\section{Discussion}

We have formalized homeostasis as the maximization of the probability of obtaining the desired observations in a MDP. 
Our two main results are: (1) Maximizing the probability of desired observation in Eq. \ref{eq:p(y,y,..,y)} is exactly solved by HMM control, Eqs. \ref{eq:policy_beta}-\ref{eq:message_beta}, while variational inference (equivalently, free energy), Eq. \ref{eq:KL_expanded}, provides a generally suboptimal policy, and (2) HMM control policy is deterministic, Eq. \ref{eq:policy_beta}, while the variational inference policy is strictly stochastic, Eq. \ref{eq:optimal_policy}.
Therefore, regardless of the specific problem considered, substantial differences between the HMM control and variational inference solutions are to be expected.
We have illustrated these generally valid conclusions with a simple example. 

We have also studied a different version of homeostasis that allows for visiting a finite range of observations. 
The maximum occupancy principle (MOP) provides a natural formalization of this setting. 
A MOP agent seeks to maximize action-observation path entropy over its lifespan (providing complexity) while at the same time staying within a safe observation range (providing structure via its embodiment). This idea may constitute a promising direction for studying the interplay between complex behavior and embodiment in natural agents.

\vspace{6pt}

\section*{Acknowledgments}
R.M.B. wrote the text and generated the results of this paper. LLMs were used only in the final phase of proofreading and in the generation of MATLAB scripts for data analysis.
This project was supported by grants funded by the Spanish Ministry of Science, Innovation and Universities (MICIU/AEI/10.13039/501100011033) and by “FEDER A way of making Europe” (ref: PID2023-146524NB), and by ICREA ACADÈMIA (2022) funded by the Catalan Institution for Research and Advanced Studies.
I thank Hideaki Shimazaki, Chris Buckley, and Manolis Mylonas for very useful discussions. I also warmly thank Hideaki Shimazaki for his hospitality during my stay at Kyoto University.

\bibliographystyle{unsrt}  
\bibliography{references}

\end{document}